\documentclass{appolb}
\usepackage{graphicx}
\usepackage[utf8]{inputenc}

\newcommand{\pc}[1]{\ensuremath{\left(#1\right)}}

\newcommand{\ev}[1]{\ensuremath{\big{<}#1\big{>}}}

\usepackage{lscape}
\usepackage{amsmath}
\usepackage{amssymb}
\usepackage{braket}
\usepackage{float}
\usepackage[utf8]{inputenc}

\begin{document}
\title{Magnetized QCD phase diagram: \\ net-baryon susceptibilities%
\thanks{Presented at {\it Excited QCD 2018}, 11-15 March 2018, Kopaonik, Serbia}%
}
\author{M\'arcio Ferreira, Pedro Costa, and Constan\c{c}a Provid\^encia, 
\address{CFisUC, Department of Physics, University of Coimbra, P-3004 - 516 
Coimbra, Portugal}
}
\maketitle
\begin{abstract}
Employing the Polyakov extended Nambu–Jona-Lasinio model, we determine the 
net-baryon number fluctuations of magnetized three-flavor quark matter. 
We show that the magnetic field changes the nature of the strange quark 
transition from crossover to first-order at low temperatures.
In fact, the strange quark undergoes multiple first-order phase transitions and 
several critical end points emerge in the phase diagram.
\end{abstract}

\section{Introduction}
The existence of a chiral critical end point (CEP) in the QCD phase diagram is 
still an open question. Its possible existence and location are important goals 
of the heavy ion collision (HIC) programs.
The effect of external magnetic fields on different regions of the phase diagram 
is very important, \textit{e.g.}, for heavy-ion collisions at very high 
energies, the early stages of the Universe and magnetized neutron stars.
 
The fluctuations of conserved quantities, such as baryon, electric, and 
strangeness charges number, play a major role in the experimental search for the 
CEP in HIC.
Experimental measurements of cumulants of net-proton (proxy for net-baryon) are
expected to carry information about the medium created by the collision 
\cite{Asakawa:2015ybt}.
The cumulants of the net-baryon number are particularly relevant as they diverge 
at the CEP \cite{Stephanov:1998dy}. 
We will study how cumulants of the net-baryon number are affected by the 
presence of magnetic fields with its consequences for the location of the CEP.
\section{Model}
The Lagrangian density of the PNJL model in the presence of an external magnetic field reads
\begin{align*}
{\cal L} &= {\bar{q}} \left[i\gamma_\mu D^{\mu}-{\hat m}_f \right ] q + 
	G_s \sum_{a=0}^8 \left [({\bar q} \lambda_ a q)^2 + ({\bar q} i\gamma_5 \lambda_a q)^2 \right ]- \frac{1}{4}F_{\mu \nu}F^{\mu \nu}\\
	&-K\left\{{\rm det} \left [{\bar q}(1+\gamma_5) q \right] + 
	{\rm det}\left [{\bar q}(1-\gamma_5)q\right]\right\} + 
	\mathcal{U}\left(\Phi,\bar\Phi;T\right).
	\label{Pnjl}
\end{align*}
The $q = (u,d,s)^T$ is the three flavor quark field with corresponding (current) 
mass matrix ${\hat m}_f= {\rm diag}_f (m_u,m_d,m_s)$. 
The the (electro)magnetic tensor is given by 
$F_{\mu \nu }=\partial_{\mu }A^{EM}_{\nu }-\partial _{\nu }A^{EM}_{\mu }$, and 
the covariant derivative $D^{\mu}=\partial^\mu - i q_f A_{EM}^{\mu}-i A^\mu$
couples the quarks to both the magnetic field $B$, {\it via} $A_{EM}^{\mu}$, and 
to the effective gluon field, {\it via} 
$A^\mu(x) = g{\cal A}^\mu_a(x)\frac{\lambda_a}{2}$, where
${\cal A}^\mu_a$ is the SU$_c(3)$ gauge field and
$q_f$ is the quark electric charge ($q_d = q_s = -q_u/2 = -e/3$).
A  static and constant magnetic field in the $z$ direction is considered, 
$A^{EM}_\mu=\delta_{\mu 2} x_1 B$. The logarithmic effective potential 
$\mathcal{U}\left(\Phi,\bar\Phi;T\right)$ \cite{Roessner:2006xn} is used, fitted
to reproduce lattice calculations ($T_0=210$ MeV). 
The divergent ultraviolet sea quark integrals are regularized by a sharp cutoff 
$\Lambda$ in three-momentum space.

The model parameters used are:
$\Lambda = 602.3$ MeV, $m_u= m_d=5.5$ MeV, $m_s=140.7$ MeV, 
$G_s^0 \Lambda^2= 1.835$, and $K \Lambda^5=12.36$ \cite{Rehberg:1995kh}.
Besides, two model variants with distinct scalar interaction coupling are 
analyzed: a constant coupling, $G_s=G_s^0$, and a magnetic field dependent 
coupling $G_s=G_s(eB)$ \cite{Ferreira:2014kpa,Ferreira:2014exa}.
The magnetic field coupling  dependence, $G_s=G_s(eB)$, reproduces the decrease 
of the chiral pseudo-critical temperature as a function of $B$ obtained in LQCD 
calculations \cite{baliJHEP2012}.

Fluctuations of conserved charges, such as the net-baryon number, provide 
important information on the effective degrees of freedom and on critical 
phenomena.
The n$^{th}$-order net-baryon susceptibility is given by
\begin{equation}
 \chi_B^n(T,\mu_B)= \frac{\partial^n\pc{P(T,\mu_B)/T^4}}{\partial(\mu_B/T)^n}.
\end{equation}
Symmetric quark matter is considered $\mu_u=\mu_d=\mu_s=\mu_q=\mu_B/3$
in the present work.
\section{Results}
\begin{figure}[!t]
	\centering
	\includegraphics[width=0.95\linewidth,angle=0.0]{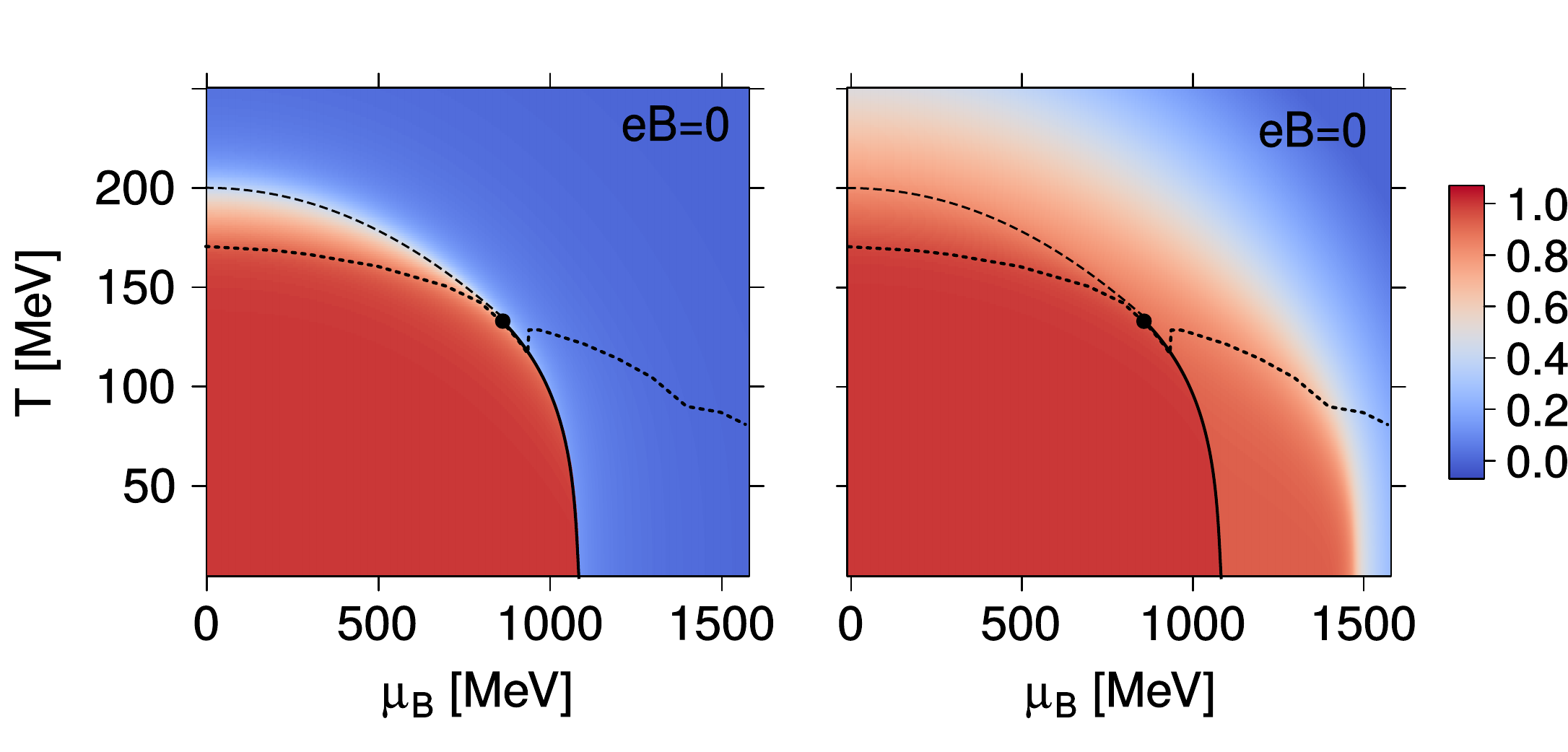}
	\caption{The (vacuum normalized) light-quark (left panel) and strange-quark 
	(right panel) condensates $\ev{q\bar{q}}(T,\mu_B)/\ev{q\bar{q}}(0,0)$. 
	The chiral first-order phase transition (solid line), the CEP (black dot), and
	both the chiral (dashed line) and deconfinement (dotted line) crossover
	boundaries are shown.}
	\label{fig:1}
\end{figure}
The quark condensates $\ev{q\bar{q}}(T,\mu_B)/\ev{q\bar{q}}(0,0)$  in the 
absence of an external magnetic field are shown in Fig. \ref{fig:1}.
While the chiral condensate (left panel) shows a crossover transition at high 
temperatures $(T>T^{\texttt{CEP}})$, it undergoes a first-order phase transition 
at lower temperatures $(T<T^{\texttt{CEP}})$. 
The first-order phase transition boundary ends up in a CEP (dot) at
$(T^{\texttt{CEP}}=133 \text{ MeV},\mu_B^{\texttt{CEP}}=862 \text{ MeV})$. 
\begin{figure}[!b]
	\centering
	\includegraphics[width=1.0\linewidth,angle=0.0]{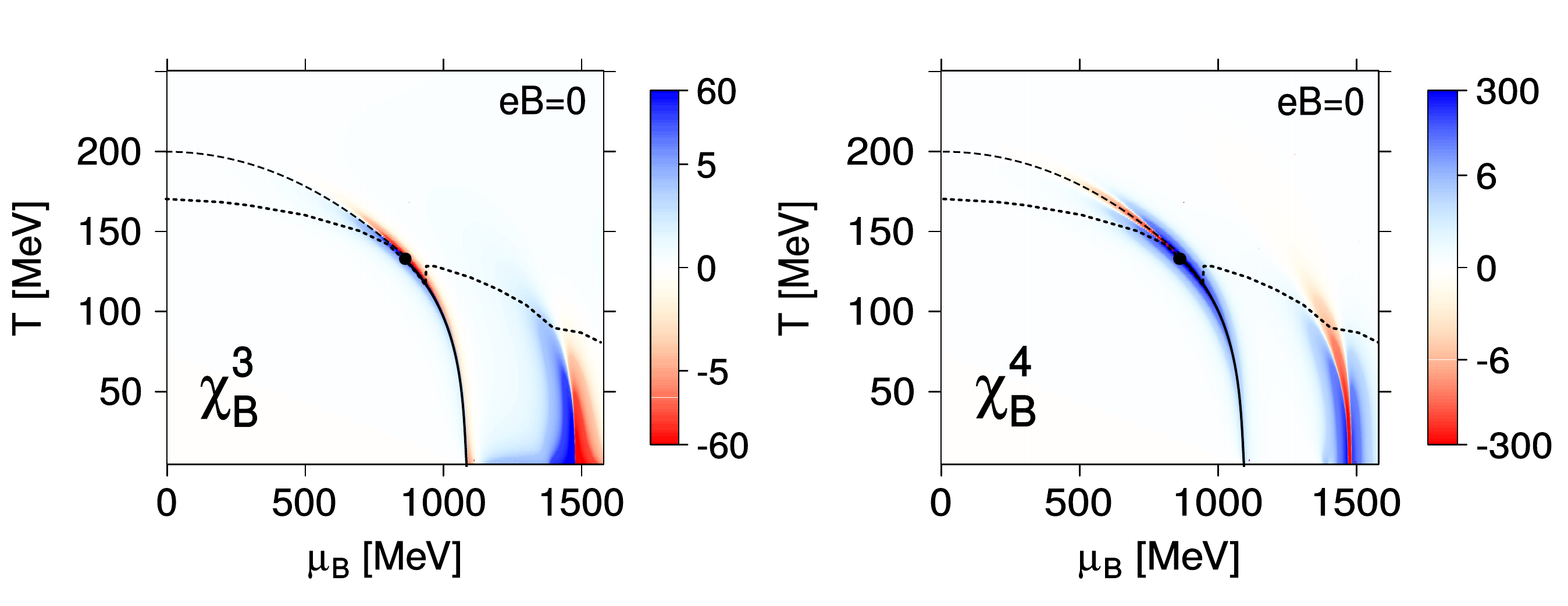}
	\caption{The $\chi^3_B$ (left panel) and $\chi^4_B$ (right panel) net-baryon 
	number susceptibilities.
	The chiral first-order phase transition (solid line), the CEP (black dot), and
	both the chiral (dashed line) and deconfinement (dotted line) crossover
	boundaries are shown.}
	\label{fig:2}
\end{figure}
Despite the strange quark condensate being discontinuous at the first-order 
chiral phase transition, its value suffers only a slight change and is still 
high (far from being approximately restored).
The decrease of the strange quark condensate, and thus the approximately 
restored phase, is attained through a crossover transition.
\begin{figure}[!t]
	\centering
	\includegraphics[width=1.0\linewidth,angle=0.0]{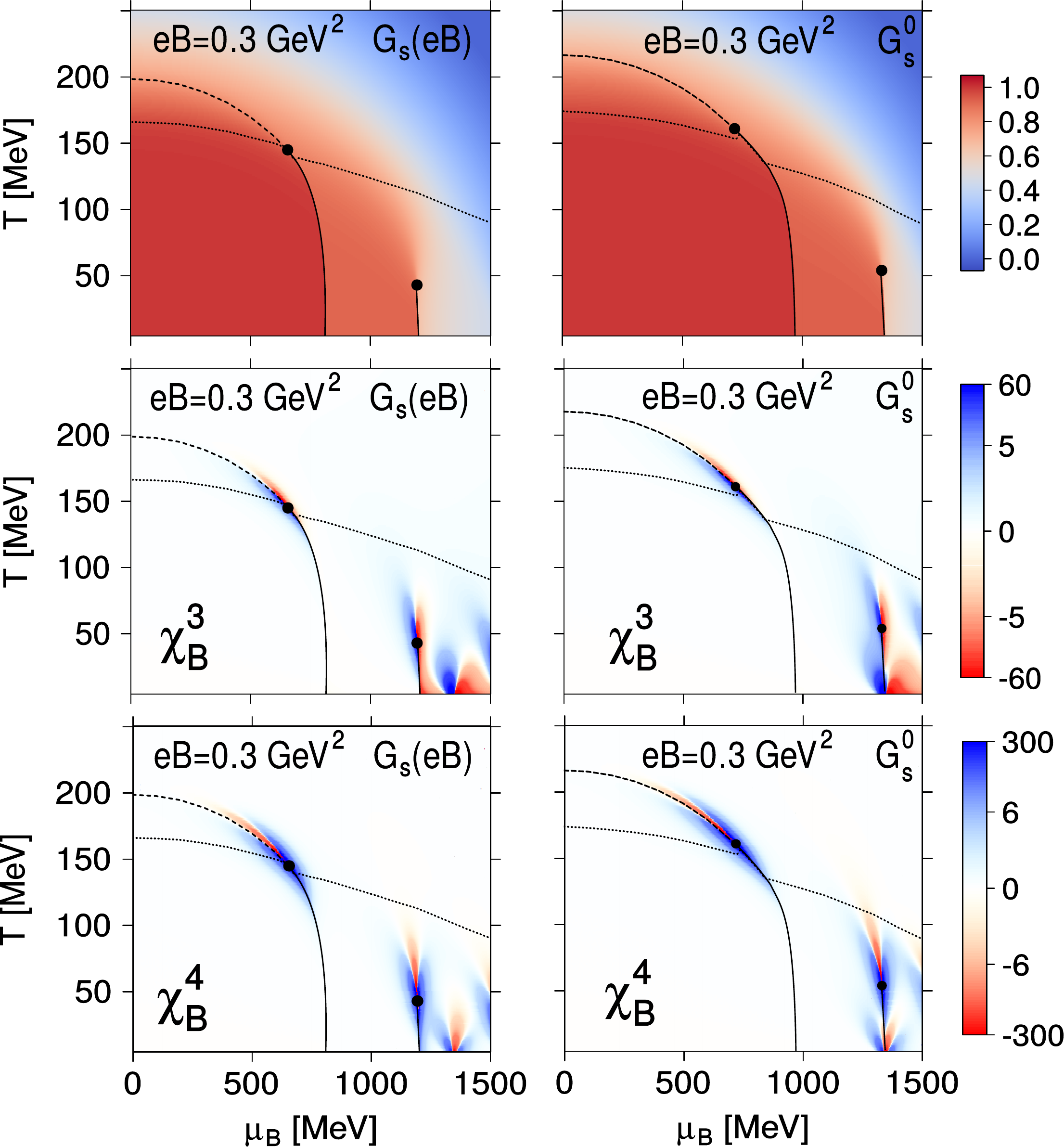}
	\caption{The (vacuum normalized) strange-quark condensate (top panel),
	the $\chi^3_B$ (middle panel) and $\chi^4_B$ (bottom panel) net-baryon number 
	susceptibilities for $G_s(eB)$ (left) and $G_s^0$ (right) models at $eB=0.3$
	GeV$^2$.
	The chiral first-order phase transition (solid line), the CEP (black dot), and
	both the chiral (dashed line) and deconfinement (dotted line) crossover 
	boundaries are shown.}
	\label{fig:3}
\end{figure}
Nevertheless, an interesting feature is seen when we look at the 
$\chi^3_B$ and $\chi^4_B$ net-baryon number susceptibilities in Fig. \ref{fig:2}.
Just as the non-monotonic dependence of the susceptibilities near the CEP, which 
signals critical phenomena, a similar structure is seen at low $T$ and 
$\mu_B\approx1500$ MeV \cite{marcio1}. 
This indicates that a slight change on the model parametrization (e.g., a 
stronger scalar coupling) might induce a first-order phase transition for the 
strange quark.
A strong external magnetic field has exactly this effect \cite{marcio2}. 
The strange quark condensate and the net-baryon number susceptibilities for both 
$G_s(eB)$ (right panel) and $G_s^0$ (left panel) models at $eB=0.3$ GeV$^2$ are 
in Fig. \ref{fig:3}.
We see that both models predict a first-order phase transition for the strange 
quark and the existence of a CEP related with the strange quark sector. Depending on the magnetic field strength, multiple phase transitions occur for 
both light and strange quarks \cite{Costa:2013zca,Ferreira:2017wtx}.
The behavior of $\chi^3_B$ and $\chi^4_B$ shows the emergence of several CEPs 
through the characteristic non-monotonic dependence, which signals the presence 
of critical behavior.\\

\vspace{-0.25cm}
{\bf Acknowledgments}:
This work was supported by ``Funda\c{c}\~ao para a Ci\^encia e a Tecnologia'', Portugal, under the projects UID/FIS/04564/2016 and POCI-01-0145-FEDER-029912, 
and under the Grants SFRH/BPD/102\
273/2014 (P.C.), and CENTRO-01-0145-FEDER-000014 (M.F.) through CENTRO2020 program.
Partial support comes from COST Action CA15213 THOR. 


\vspace{-0.25cm}
\begin{thebibliography}{99}


\bibitem{Asakawa:2015ybt} 
  M.~Asakawa and M.~Kitazawa,
  Prog.\ Part.\ Nucl.\ Phys.\  {\bf 90}, 299 (2016),
  [arXiv:1512.05038 [nucl-th]].
  
\bibitem{Stephanov:1998dy} 
  M.~A.~Stephanov, K.~Rajagopal and E.~V.~Shuryak,
  Phys.\ Rev.\ Lett.\  {\bf 81}, 4816 (1998),
  [hep-ph/9806219].
  
\bibitem{Roessner:2006xn} 
  S.~Roessner, C.~Ratti and W.~Weise,
  Phys.\ Rev.\ D {\bf 75}, 034007 (2007),
  [hep-ph/0609281].

\bibitem{Rehberg:1995kh} 
  P.~Rehberg, S.~P.~Klevansky and J.~Hufner,
  Phys.\ Rev.\ C {\bf 53}, 410 (1996),
  [hep-ph/9506436].

\bibitem{Ferreira:2014kpa} 
  M.~Ferreira, P.~Costa, O.~Louren\c{c}o, T.~Frederico, and C.~Provid\^encia,
  Phys.\ Rev.\ D {\bf 89}, no. 11, 116011 (2014),
  [arXiv:1404.5577 [hep-ph]].

\bibitem{Ferreira:2014exa} 
  M.~Ferreira, P.~Costa and C.~Provid\^encia,
  Phys.\ Rev.\ D {\bf 90}, no. 1, 016012 (2014),
  [arXiv:1406.3608 [hep-ph]].

\bibitem{baliJHEP2012}
	G. S. Bali, F. Bruckmann, G. Endr\"odi, Z. Fodor, S. D. Katz, S. Krieg, 
  A. Sch\"afer, and K. K. Szab\'o,
  J. High Energy Phys. {\bf 1202}, 044 (2012),
  [arXiv:1111.4956 [hep-lat]].
	
\bibitem{marcio1} 
	M.~Ferreira, P.~Costa and C.~Provid\^encia,
  Phys.\ Rev.\ D {\bf 98}, no. 3, 034003 (2018),
  [arXiv:1806.05758 [hep-ph]].

\bibitem{marcio2} 
	M.~Ferreira, P.~Costa and C.~Provid\^encia,
  Phys.\ Rev.\ D {\bf 98}, no. 3, 034006 (2018),
  [arXiv:1806.05757 [hep-ph]].

\bibitem{Costa:2013zca} 
  P.~Costa, M.~Ferreira, H.~Hansen, D.~P.~Menezes and C.~Provid\^encia,
  Phys.\ Rev.\ D {\bf 89}, no. 5, 056013 (2014),
  [arXiv:1307.7894 [hep-ph]].

\bibitem{Ferreira:2017wtx} 
  M.~Ferreira, P.~Costa and C.~Provid\^encia,
  Phys.\ Rev.\ D {\bf 97}, no. 1, 014014 (2018),
  [arXiv:1712.08378 [hep-ph]].

\end{thebibliography}
\end{document}